\documentclass[10pt, onecolumn, showpacs]{revtex4}
\usepackage{amsmath}
\usepackage[dvips]{graphicx}
\usepackage{tabularx}
\usepackage{color}

\begin{document}

\title{Segregation pattern reorientation of granular mixture on horizontally oscillating tray}
\author{Masashi Fujii}
\email{mfujii0123@hiroshima-u.ac.jp}
\affiliation{
	Department of Mathematical and Life Sciences, Hiroshima University, Kagamiyama, Higashi-Hiroshima, 739-8526,  Japan.\\
}
\author{Akinori Awazu}
\affiliation{
	Department of Mathematical and Life Sciences, Hiroshima University, Kagamiyama, Higashi-Hiroshima, 739-8526,  Japan.\\
}
\author{Hiraku Nishimori}
\affiliation{
	Department of Mathematical and Life Sciences, Hiroshima University, Kagamiyama, Higashi-Hiroshima, 739-8526,  Japan.\\
}

\begin{abstract}

	Reorientation of the segregation pattern of a binary granular mixture on a two-dimensional horizontally oscillating tray is numerically realized.
	The mixture consists of large-and-heavy particles and small-and-light particles,
	the segregation pattern of which shows a transition between a striped pattern perpendicular to the oscillation and that parallel to the oscillating direction according to the change of area-fractions of two types of particles.
	The transition mechanism is discussed on the basis of a simplified 1-dimensional dynamics.
	
\end{abstract}

\maketitle

%%%%%%%%%%%					Intoroduction				%%%%%%%%%%%
%\section{Introduction}
	Granular materials exhibit various complex behaviors, and their characteristic dynamics have been studied extensively,
	for instance, the crystallization of granular materials\cite{reis2006a} or the pattern formation of granular mixtures on a horizontally oscillating tray \cite{ciamarra2007, reis2004, reis2006b, ehrhardt2005},
	in a horizontally rotating drum \cite{choo1997, zik1994, awazu2000, stavans1998, hill1997, krzyzewski2008}, and in a vertically oscillating box \cite{reis2007, lipowski2002}.
	However, thus far, the research on the basic physics of the granular dynamics as well as macroscopic phenomenology has been insufficient.
	For example, previous studies on the patten formation of a granular mixture on a horizontally oscillating tray \cite{ciamarra2007, reis2004, reis2006b, ehrhardt2005}
	have demonstrated that different types of granular materials segregate to form a stripe pattern that extends perpendicular to the oscillating direction, while no segregated pattern has been observed parallel to the oscillating direction.
	The potential possibility of the emergence of the latter segregation pattern has been an intriguing issue in this field of study.
	For a wider range of soft materials, a recent numerical study on a colloidal mixture found stripe patterns that were both parallel and perpendicular to the oscillating electric field \cite{wysocki2009}.
	In this study, we use a numerical model and show that transition between steady perpendicular and steady parallel stripe patterns can be observed in granular mixture as well,
	for a suitable set of parameters and a suitable combination of the area fractions of the two types of constituting particles.

%%%%%%%%%%%					Model				%%%%%%%%%%%
%\section{Model}
	We employ the distinct element method (DEM) \cite{ciamarra2007, ehrhardt2005} and simulate the behaviors of the granular mixture on a two-dimensional horizontally oscillating tray.
	This mixture consists of two types of particles:
	$i)$ the particle with a large grain size and mass, which we refer to as the $L$-particle, and
	$ii)$ the particle with a small grain size and mass,  which we refer to as the $S$-particle (see Table \ref{tab:table1}).
	The $i$-th particle has type-dependent mass ($m_i^{type}$) and size ($D_i^{type}$) $(type = L\ \mbox{or}\ S)$, and its position is denoted by a two-dimensional vector, ${\bf{r}}_i(=(x_i,y_i))$.
	Each particle is subjected to viscous friction, which is proportional to the velocity difference between the particle and the tray with the friction coefficient $\mu$ \cite{ciamarra2007, ehrhardt2005}.
	If the $i$-th and $j$-th particles overlap, the interaction is indicated by a linear elastic force with viscous dissipation, depending on the combination of the types of particles;
	the elastic coefficient and the coefficient of dissipation are denoted as $k$ and $\gamma_{i,j}$, respectively.
	Here, $\gamma_{i,j}$ is derived from the coefficient of restitution $e^{i,j}$, the elastic coefficient $k$, and the reduced mass of particles as follows.  
	\begin{align}
		\gamma_{i,j} = -\frac{2\log e_{i,j}\sqrt{k\frac{ m_i m_j}{m_i+m_j}}}{\sqrt{\pi^2+(\log e_{i,j})^2}}.
		\label{eq:equation1}
	\end{align}
	The dynamic equation of the $i$-th particle is summarized as follows.
	\begin{align}
		m_i^{type} \ddot {\bf{r}}_i&=\sum_{j\neq i}{\bf{f}}_{i,j}-\mu (\dot {\bf{r}}_i - \dot {\bf{r}}_{\mbox{tray}}),
		\label{eq:equation2}
	\end{align}
	\begin{align}
		{\bf{f}}_{i,j}&=\left\{
			\begin{array}{l}
				k\frac{\delta {\bf r}_{i,j}}{|\delta {\bf r}_{i,j}| }\left[\frac{D_i^{type}+D_j^{type}}{2}-|\delta {\bf{r}}_{i,j}|\right]-\gamma_{i,j}\dot{\delta {\bf r}}_{i,j}\\
																	\hspace{14mm}\left(\mbox{for}~ |\delta {\bf r}| \le {(D_i^{type}+D_j^{type})}/{2}\right)\\
				0													\hspace{12mm}\left(\mbox{for}~ |\delta {\bf r}| >   {(D_i^{type}+D_j^{type})}/{2}\right),
			\end{array}
		\right. 
		\label{eq:equation3}
	\end{align}
	\begin{align}
		\delta {\bf r}_{i,j}&={\bf r}_i-{\bf r}_j,
		\label{eq:equation4}
	\end{align}
	where the velocity of tray is given as
	\begin{align}
		\dot {\bf{r}}_{\mbox{tray}} = (\dot x_{\mbox{tray}}, \dot y_{\mbox{tray}}) =
			\left(
				2\pi A\nu \sin (2\pi \nu t),0
			\right).
		\label{eq:equation5}
	\end{align}
	\begin{table}[tb]
		\caption{Parameter values considered in numerical simulation for pattern transformation.}
		\begin{center}
			\begin{tabularx}{85mm}{|X||c|c|} \hline
				property				&$L$-value	&$S$-value\\ \hline
				$m^{type}$ : 	Mass	&$m^L=1 ~\mathrm{g}$	&$m^S=0.025 ~\mathrm{g}$\\ \hline
				$D^{type}$ : 	Size		&$D^L=1 ~\mathrm{cm}$	&$D^S=0.5 ~\mathrm{cm}$\\ \hline
			\end{tabularx}\\[-0.5mm]
			\begin{tabularx}{85mm}{|lX||c|} \hline
				$e^{type_i,type_j}$ : 	&(type of $i$, type of $j$) 	&\\
						&$(L,L)$				&$e^{L,L}=0.2$\\
						&$(L, S)$ or (S,L)		&$e^{L,S}=0.5$\\
						&$(S, S)$				&$e^{S,S}=0.9$\\ \hline
			\end{tabularx}\\[-0.5mm]
			\begin{tabularx}{85mm}{|X||c|} \hline
				$\mu$ : 			Friction coefficient				&$1 ~\mathrm{g~s^{-1}}$\\ \hline
				$k$ : 			Elastic coefficient				&$1.0\times 10^4 ~\mathrm{g~cm^2~s^{-2}}$\\ \hline
				$A$ :				Amplitude of oscillation of tray	&$10~\mathrm{cm}$\\ \hline
				$\nu$ : 			Frequency of oscillation of tray	&$2~\mathrm{Hz}$\\ \hline
			\end{tabularx}
		\end{center}
		\label{tab:table1}
	\end{table}
	Note that the rotation of the particle is not considered.
	The values of parameters are summarized in Table \ref{tab:table1},
	most of which are set same or close to those introduced in previous related studies \cite{ciamarra2007, ehrhardt2005}.
	As one of exceptions, the elastic coefficient $k$ is set smaller than that in previous study, meaning that the particles are not so rigid, but the collision time derived from the elastic coefficient remains small compared with the period of oscillation of tray.
	In addition, for the simpleness, we consider only the case where the elastic coefficients of $L$- and $S$-particles are identical.
	Another exception is the large value of $\mu$ assuming that particles are more strongly affected by the viscous friction arising from their relative velocity to the tray than previous studies.
	This is to explore the potentially new aspect of the segregation dynamics on a horizontally oscillating tray.
	Here, the random fluctuation in the system is neglected.
	Systematic calculations to verify the effect of viscous friction on segregation will be presented later.
	
	We perform simulations using several combinations of $N_L$ and $N_S$, $i.e.$ the number of $L$- and $S$-particles, respectively.
	In addition, we define the area fractions of $L$- and $S$-particles as $\rho_L\equiv N_L\pi (D^L)^2/(4W^2)$ and $\rho_S\equiv N_S\pi (D^S)^2/(4W^2)$, respectively.
	The width of the system is $W=40~\mathrm{cm}$ in both oscillational and lateral directions, and the periodic boundary conditions are set in both directions.
	Each run starts from $t=0$, when the particles are randomly distributed on a tray without overlap, and ends at $t=200$.

%%%%%%%%%%%					Results				%%%%%%%%%%%
%\section{Result}
	\begin{figure}[tb]
		\begin{center}
			(a)
			\begin{minipage}[t][][b]{30mm}
				\includegraphics[width=30mm]{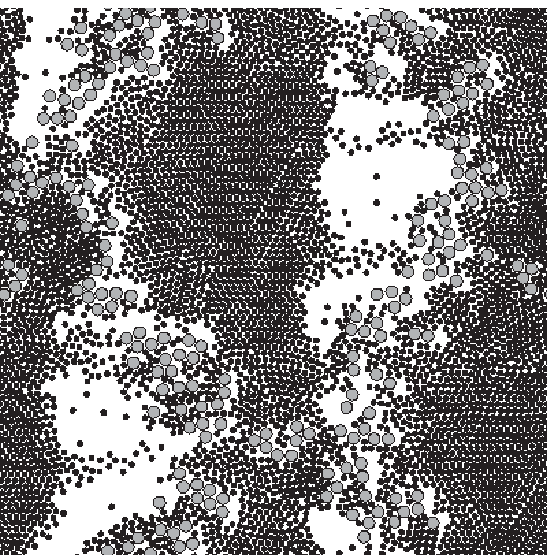}
			\end{minipage}
			(b)
			\begin{minipage}[t][][b]{30mm}
				\includegraphics[width=30mm]{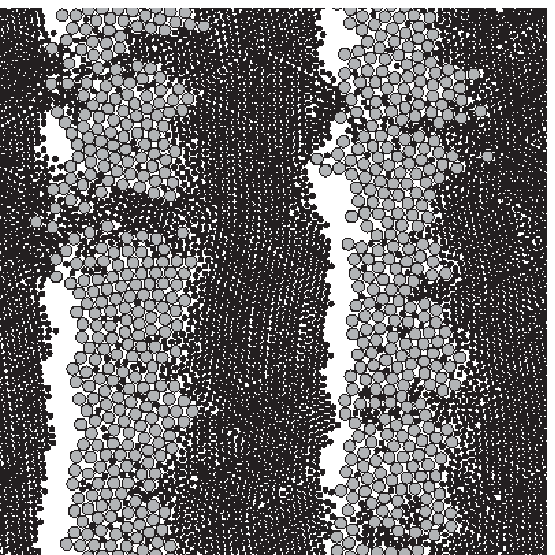}
			\end{minipage}\\
			(c)
			\begin{minipage}[t][][b]{30mm}
				\includegraphics[width=30mm]{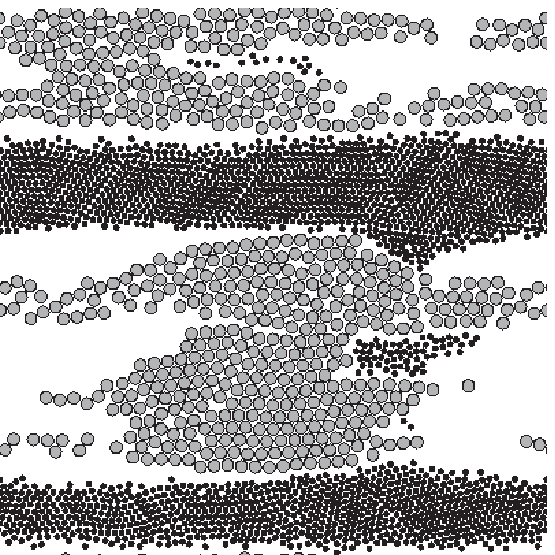}
			\end{minipage}
			(d)
			\begin{minipage}[t][][b]{30mm}
				\includegraphics[width=30mm]{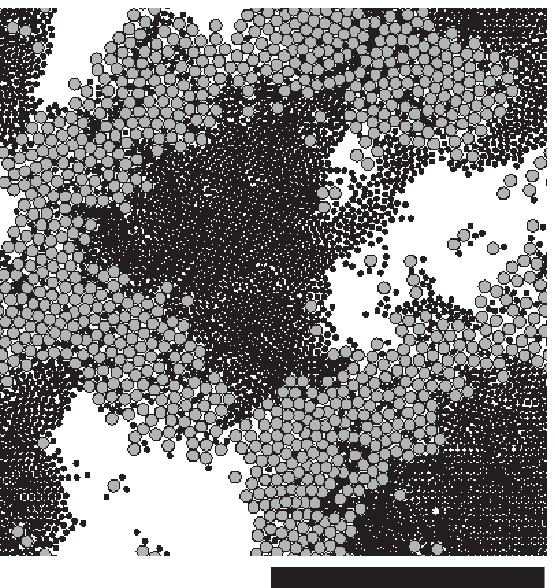}
			\end{minipage}
			\begin{flushleft}
				\hspace{12mm}{\Large $\longleftrightarrow$}\hspace{2mm}oscillating direction
			\end{flushleft}
			\caption{
				Typical steady states of particle mixtures:
				(a) non-stripe pattern ($\rho_L=0.1, \rho_S=0.65$).
				(b) stripe pattern perpendicular to oscillating direction ($\rho_L=0.30, \rho_S=0.55$),
				(c) stripe pattern parallel to oscillating direction ($\rho_L=0.30, \rho_S=0.25$), and
				(d) crossed-striped pattern ($\rho_L=0.40, \rho_S=0.40$).
				The scale bar denotes $20\mathrm{cm}$.
			}
			\label{fig:figure1}
		\end{center}
	\end{figure}
	Figure \ref{fig:figure1} shows typical patterns emerging as steady states of the binary granular particle mixtures:
	(a) non-stripe pattern,
	(b) stripe pattern perpendicular to the oscillating direction (hereafter, we refer to this pattern as the perpendicular-striped pattern),
	(c) stripe pattern parallel to the oscillating direction (hereafter, we refer to this pattern as the parallel-striped pattern), and
	(d) crossed-striped pattern.
	
	To distinguish these four patterns quantitively, we define a temporal order parameter $\Phi(t)$.
	Specifically, the system is separated into $M\times M~(M=40)$ square cells and a quantity,
	\begin{align}
		\sigma_{i,j}(t)=N_{i,j}^L(t)\cdot (D^L)^2-N_{i,j}^S(t)\cdot (D^S)^2,
		\label{eq:equation6}
	\end{align}
	is allocated to each cell $(i,j)$, where $N_{i,j}^L(t)$ and $N_{i,j}^S(t)$ are, respectively, the number of $L$- and $S$-particles in cell $(i, j)$ at $t$.
	Then, the specific form of the temporal order parameter $\Phi(t)$ is given as
	\begin{align}
		\Phi(t)=&
		 \sum_j \sum_k\frac{\sum_i [\sigma_{i,j}(t)\sigma_{i+k, j}(t)]}{M^2 \sum_i [\sigma_{i,j}(t)^2]}\nonumber\\
		&-\sum_i \sum_k\frac{\sum_j [\sigma_{i,j}(t)\sigma_{i, j+k}(t)]}{M^2 \sum_j [\sigma_{i,j}(t)^2]}.
		\label{eq:equation7}
	\end{align}
	The first and second terms of the right-hand-side of Eq. (\ref{eq:equation7}) quantify the spatial correlations along the direction parallel and perpendicular to the oscillating direction, respectively.
	In simple terms, $\Phi(t)$ approaches 1 when the parallel-striped pattern is dominant, whereas it approaches -1 when the perpendicular-striped pattern is dominant.
	
\begin{figure}[tb]
		\begin{center}
			\includegraphics[width=80mm]{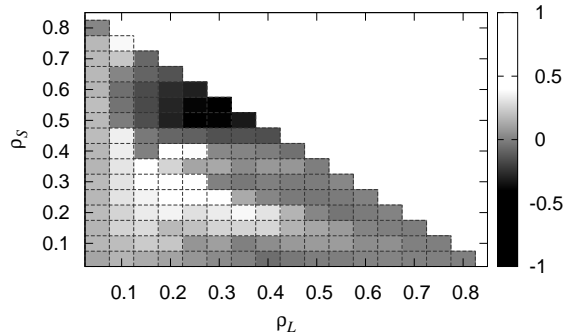}
			\caption{
				Relationship between the order parameter $\Phi_{av}$ and the area fraction of $L$-particles (horizontal axis) and $S$-particles (vertical axis), for $A=10$ and $\nu=2$.
				A bright color implies that $\Phi_{av}$ is close to 1, indicating the emergence of a parallel-striped pattern,
				and a dark color implies that $\Phi_{av}$ is close to -1, indicating the emergence of a perpendicular-striped pattern.
			}
			\label{fig:figure2}
		\end{center}
	\end{figure}

	Simulations are systematically performed by varying the combination of $\rho_L \in\{ 0.05, 0.1, 0.15, \cdot\cdot\cdot 0.8 \}$ and $\rho_S \in\{ 0.05, 0.1, 0.15, \cdot\cdot\cdot 0.8 \}$, maintaining $\rho_S + \rho_L \le 0.85$.
	The relationship between $(\rho_L, \rho_S)$ and $\Phi_{av}$, with $\Phi(t)$ averaged over the last 20 periods of oscillation, for $A=10$ and $\nu=2$ is shown in Fig. \ref{fig:figure2}.
	As shown in the figure, to observe this relationship, we consider four typical regions:
	(a) $\rho_L \le 0.1$ or $\rho_S \le 0.1$, where $\Phi_{ave}$ is close to zero;
	(b) $\rho_L+\rho_S \le 0.85$, $\rho_L \ge 0.15$, and $\rho_S \ge 0.45$, where $\Phi_{av}$ is close to -1;
	(c) $\rho_L+\rho_S \le 0.55$, $0.15 \le \rho_L \le 0.40$, and $0.15\le \rho_S \le 0.35$, where $\Phi_{av}$ is close to 1; and
	(d) $0.55 \le \rho_L+\rho_S \le 0.85$ and $0.20 \le \rho_S \le 0.40$, where $\Phi_{ave}$ is close to zero.
	In region (a), particles tend not to form any stripe pattern, and in regions (b), (c), and (d), particles tend to form the perpendicular-striped, the parallel-striped, and the crossed-striped patterns, respectively.
	
	\begin{figure}[tb]
		\begin{center}
			\includegraphics[width=80mm]{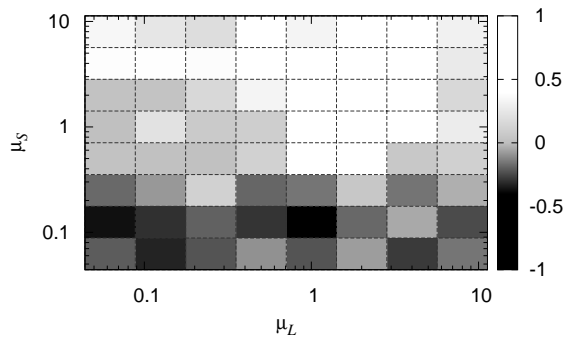}
			\caption{
				Relationship between the order parameter $\Phi_{av}$ and the friction coefficients of $L$-particles (horizontal axis, log scale) and $S$-particles (vertical axis, log scale), for $\rho_L = 0.25$, $\rho_S = 0.3$, $A=10$ and $\nu=2$.
				A bright color implies that $\Phi_{av}$ is close to 1, indicating the emergence of a parallel-striped pattern,
				and a dark color implies that $\Phi_{av}$ is close to -1, indicating the emergence of a perpendicular-striped pattern.
			}
			\label{fig:figure3}
		\end{center}
	\end{figure}
	
	\begin{figure}[tb]
		\begin{center}
			\includegraphics[width=80mm]{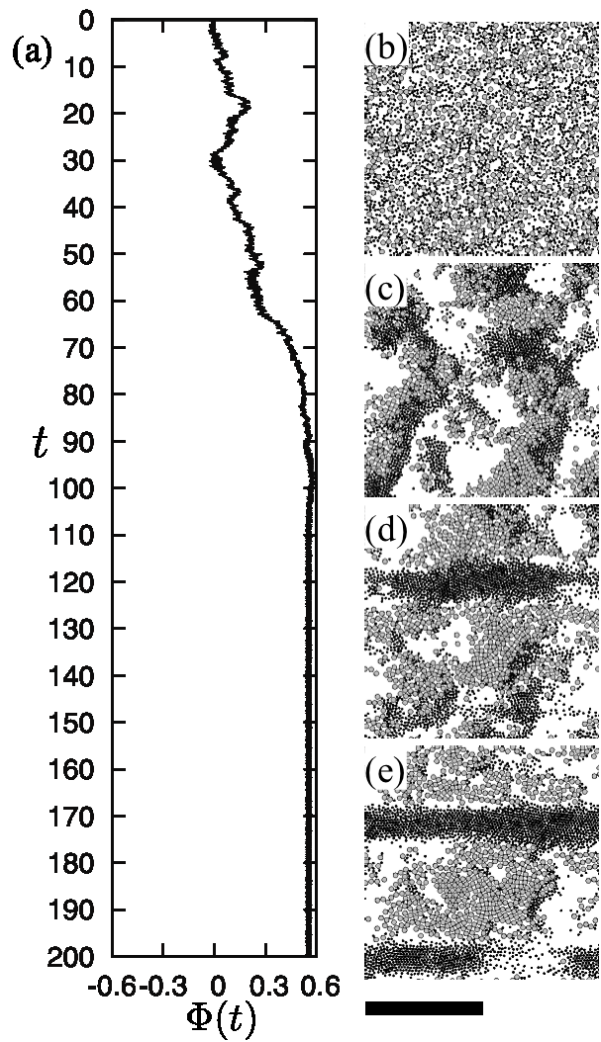}
			\caption{
				(a)Time evolution of the order parameter, $\Phi(t)$, for $\rho_L=0.30$ and $\rho_S=0.25$, and
				the corresponding simulation snapshots at (b) $t=0$, (c) 30, (d) 60 and (e) 70.
				The scale bar denotes $20\mathrm{cm}$ for (b)-(d).
			}
			\label{fig:figure4}
		\end{center}
	\end{figure}
	
	It should be noted that the parallel-striped pattern obtained in region (c) has not been realized in previous studies on the horizontally oscillating granular mixture,
	wherein much lower viscous friction ($\mu\sim 0.1$) has been used.
	To verify the effect of  high viscosity friction, $\mu=1$, considered in the above calculations, as shown in Fig. \ref{fig:figure3},
	we calculate the order parameter in steady state with a varying the combination of friction coefficient of $L$-particles and $S$-particles keeping $\rho_L = 0.25$, $\rho_S = 0.3$, $A=10$ and $\nu=2$.
	The case of $\mu_L=\mu_S=1$ in this figure corresponds to the case of $\rho_L=0.25, \rho_S=0.3$ in Fig. \ref{fig:figure2}.
	As roughly shown in Fig. \ref{fig:figure3}, particles tend to form the parallel-striped pattern in the high-viscous-friction range, and the perpendicular-striped pattern in the low-viscous-friction range.
	This is the reason why we choose the set, $\mu_L=\mu_S=1$, of viscous friction coefficients in this study with which set we can see the present characteristic segregation dynamics approaching the parallel-striped pattern.
	Now, we focus on the time evolution of $\Phi(t)$ in region (c).
	Figure \ref{fig:figure4}(b)-\ref{fig:figure4}(e) shows simulation snapshots at $t=0, 30, 60$, and 70, respectively, at $(\rho_L, \rho_S)=(0.3, 0.25)$;
	there, the gray particles correspond to $L$-particles and the black particles correspond to $S$-particles.
	At the early stage of the simulation, $0 < t \le 30$, particles temporarily form the perpendicular-striped-like pattern as shown in Fig. \ref{fig:figure4}(c), where the value of $\Phi(t)$ is kept around zero.
	In this stage, the perpendicular-striped pattern arises, but alignments of stripes are not so distinct.
	In the next stage $t = 30$ until $t=90$, the value of $\Phi(t)$ gradually increases, accordingly, the striped-pattern is largely deformed to form clusters of $L$-particles.
	For example, after $t=60$, the perpendicular stripe is hardly recognizable as seen in Fig. \ref{fig:figure4}(d),
	instead, an wedge of horizontally extending cluster of $S$-particles is seen to penetrate into the the clusters of $L$-particles, which serves as a seed of parallel-striped pattern.
	Subsequently, another horizontal wedge of $S$-particles begins to collide the clusters of $L$-particles to form the 2nd parallel stripe as seen in Fig \ref{fig:figure4}(e) for $t=70$.
	Around $t=90$, $\Phi(t)$ reaches a saturation value $\Phi (t)\simeq 0.55$, before which time the parallel-striped pattern has been established,
	in that situation, the perpendicular movement of particles is restricted by the strong friction force acting between the particles and the horizontally oscillating tray.
	Hence, the movements of $L$- and $S$-particles get synchronized,
	resulting in a sharp decrease in the frequency of particle collisions.
	In this way, the deformation of perpendicular stripes, which are formed once in the system, irrespective of the final pattern, plays a key role in the selection of the final pattern.
	In particular, considering the far more active movement of $S$-particles than $L$-particles,
	that causes the deformation of the perpendicular alignment of $L$-particles,
	the stability of stripes of $L$-particles against the collisions of the cluster of $S$-particles seems essential for the selection of final pattern selection; the perpendicular-striped pattern or the parallel-striped pattern.
	
\begin{figure}[tb]
		\begin{center}
			\includegraphics[width=60mm]{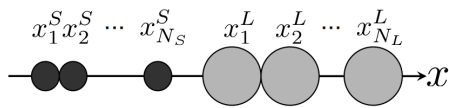}
			\caption{
				Illustration of the system of a one-dimensional model.
				The small black circle represents the $S$-particle and the large gray circle represents the $L$-particle. 
			}
			\label{fig:figure5}
		\end{center}
	\end{figure}

	Accordingly, to investigate the stability of $L$-particle stripes (cluster) in a simplified setup,
	we consider a one-dimensional (1D) granular system, in which particles arraying in the $x$ (oscillating) direction share the central axis, as shown in Fig. \ref{fig:figure5}; the $L$- and $S$-particles are kept unmixed.
	This is the 1D expression of a-pair-of-perpendicular-striped pattern realized in the 2D system.
	The dynamics and external force are kept the same as those in 2D system, where the boundary condition is periodic.
	Note that because of the dimensionality, the sequence of particles is preserved in the 1D simulation.
	In this situation, we do not focus on the stability of the initial sequence of the $L$- and $S$-particles but on the stability of the agglutinated state of the $L$-particles.
	This agglutinated state of the $L$-particles corresponds to a rigid structure of individual perpendicular stripes.

	To quantify the agglutination degree of the $L$-particles in the steady state of the system,
	we measure the mean distance between neighboring $L$-particles,
	\begin{align}
		x_{ave}^L=\frac{1}{\tau}\int_{\tau}^{T}\frac{\left[x_{N_L}^L(t) - x_1^L(t)\right]\mod(W)}{N_L-1} dt,
		\label{eq:equation8}
	\end{align}
	where $W(=20)$ is the total length of the system.
	
	\begin{figure}[bt]
		\begin{center}
			\vspace{-5mm}	\includegraphics[width=80mm]{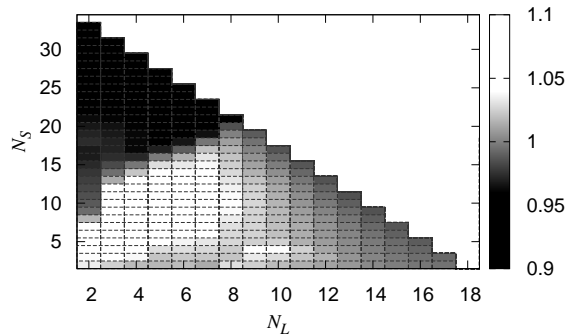}
			\caption{
				Relationships among $N_L$ (horizontal axis), $N_S$ (vertical axis) and $x_{ave}^L$, and MSD.
				The very dark color indicates that the value of $x_{ave}$ is smaller than 1.
				The lighter color indicates that the value of $x_{ave}^L$ is larger than 1.
			}
			\label{fig:figure6}
		\end{center}
	\end{figure}
	The relationship among $N_L$, $N_S$, and $x_{ave}^L$, obtained by the simulation, is shown in Fig. \ref{fig:figure6}.
	The very dark color indicates that the value of $x_{ave}$ is smaller than 1,
	the lighter color indicates that the value of $x_{ave}^L$ is larger than 1.
	In this figure, we see three typical regions.
	In the cases where $N_S < 15$ and $N_L < 9$, the mean distance between the $L$-particles is larger than 1;
	in other words, the $L$-particles cannot be kept agglutinated in the field.
	It should be noted that, in this region, $L$- and $S$-particles frequently collide at their boundaries.
	In contrast, in the case where $N_S \ge 20$, the obtained values of $x_{ave}^L$ are smaller than 1, implying that the agglutinated state of $L$-particles is stably maintained.
	In this case, the frequency of the collisions between the $S$- and $L$-particles is much lower than that in the above case.
	In the case where $N_L>10$, $x_{ave}^L$ takes intermediate values close to 1.
	These results suggest that the destabilization of the agglutinated state of $L$-particles in 1D simulation is realized for small spatial densities of $L$- and $S$-particles;
	this destabilization of the agglutinated state roughly corresponds to the unstable condition of perpendicular stripes of $L$-particles in 2D simulation shown in Fig. \ref{fig:figure2}.
	On the other hand, in the high-density region of the 1D and 2D simulations, particularly in the high-density regime of $S$-particles, the agglutinated state of the $L$-particles is maintained stable.
	In this way, the formation of the parallel-striped pattern is caused by the destabilization of the $L$-particle stripes.
	This destabilization requires frequent collisions of $S$-particles, that is, the momentum transfer of $S$-particles, to the $L$-particles striped structure.
	This momentum transfer is realized when $S$-particles move actively in the field for a proper combination of the area fractions of $S$- and $L$-particles.
	The two destabilization processes are to be investigated in greater detail in future research.

%%%%%%%%%%%					Conclusion and Discussion				%%%%%%%%%%%
%\section{conclusion and discussion}
	We numerically investigated the pattern formation of granular mixtures on a horizontally oscillating tray
	and found that a stripe pattern parallel to the oscillating direction is formed for small area fractions of particles.
	Focusing on the formation process of the parallel-striped pattern,
	the particles temporarily form a stripe pattern perpendicular to the oscillating direction of the tray in the early stage,
	hence, the perpendicular stripes of large and heavy particles are broken by repeated collisions with actively moving small and light particles, finally resulting in the formation and stabilization of the parallel-striped pattern.
	
	It is noticeable that Alam {\it et al.} (2008) \cite{alam2008} reported the emergence of parallel-striped patterns in the horizontally oscillating particulate suspension.
	In the present study, we did not consider the fluid motion, however, these two systems potentially belong to the same universality class in terms of instability of the striped pattern formation parallel to horizontally oscillating external force.
	The common backgrounding mechanism for the emergence of this type of patterns should be explored as the next issue.
	Also the more systematic study for the dependency of the final segregation pattern on the elastic constants of $L$- and $S$-particle remains to be made.
	
%\section*{acknowledgment}
	This work is supported by the Research Fellowship of the Japan Society for the Promotion of Science for Young Scientists to MF,
	by Grant-in-And for Scientific Research (C) 22540391 to HN,
	and by the Global COE Program G14 (Formation and Development of Mathematical Sciences Based on Modeling and Analysis)
	of the Ministry of Education, Culture, Sports, Science and Technology of Japan to HN.

\end{document}